# Topological magnetoelectric effects in microwave far-field radiation

M. Berezin, E.O. Kamenetskii, and R. Shavit

Microwave Magnetic Laboratory,
Department of Electrical and Computer Engineering,
Ben Gurion University of the Negev, Beer Sheva, Israel

June 25, 2015

**Abstract**
Similar to electromagnetism, described by the Maxwell equations, the physics of magnetoelectric (ME) phenomena deals with the fundamental problem of the relationship between electric and magnetic fields. Despite a formal resemblance between the two notions, they concern effects of different natures. In general, ME-coupling effects manifest in numerous macroscopic phenomena in solids with space and time symmetry breakings. Recently it was shown that the near fields in the proximity of a small ferrite particle with magnetic-dipolar-mode (MDM) oscillations have the space and time symmetry breakings and topological properties of these fields are different from topological properties of the free-space electromagnetic (EM) fields. Such MDM-originated fields – called magnetoelectric (ME) fields – carry both spin and orbital angular momentums. They are characterized by power-flow vortices and non-zero helicity. In this paper, we report on observation of the topological ME effects in far-field microwave radiation based on a small microwave antenna with a MDM ferrite resonator. We show that the microwave far-field radiation can be manifested with a torsion structure where an angle between the electric and magnetic field vectors varies. We discuss the question on observation of the regions of localized ME energy in far-field microwave radiation.

## I. INTRODUCTION

In a large variety of light-matter-interaction phenomena, novel engineered fields are considered as very attractive instruments for study different enantiomeric structures. Recently, significant interest has been aroused by a rediscovered measure of helicity in optical radiation – commonly termed optical chirality density – based on the Lipkin's "zilch" for the fields [1]. The optical chirality density is defined as [1 – 4]:

$$\chi = \frac{\varepsilon_0}{2} \vec{E} \cdot \nabla \times \vec{E} + \frac{1}{2\mu_0} \vec{B} \cdot \nabla \times \vec{B}. \tag{1}$$

This is a time-even, parity-odd pseudoscalar parameter. Lipkin showed [1], that the chirality density is zero for a linearly polarized plane wave. However, for a circularly polarized wave, Eq. (1) gives a nonvanishing quantity. Moreover, for right- and left-circularly polarized waves one has opposite signs of parameter $\chi$. The optical chirality density is related to the corresponding chirality flow via the differential conservation law:

$$\frac{\partial \chi}{\partial t} + \vec{\nabla} \cdot \vec{f} = 0, \tag{2}$$

where

$$\vec{f} = \frac{\varepsilon_0 c^2}{2} \left[ \vec{E} \times (\nabla \times \vec{B}) - \vec{B} \times (\nabla \times \vec{E}) \right]. \quad (3)$$

For time-harmonic fields (with the field time dependence $e^{i\omega t}$), the time-averaged optical chirality density is calculated as [1 – 4]

$$\chi = \frac{\omega \varepsilon_0}{2} \text{Im}(\vec{E}^* \cdot \vec{B}), \quad (4)$$

where vectors $\vec{E}$ and $\vec{B}$ are complex amplitudes of the electric and magnetic fields.

The effect of optical chirality was applied recently for experimental detection and characterization of biomolecules [5]. The chiral fields were generated by the optical excitation of plasmonic planar chiral structures. Excitation of molecules is considered as a product of the parameter of optical chirality with the inherent enantiometric properties of the material. In experiments [5], the evanescent near-field modes of plasmonic oscillations are involved. In continuation of these studies, a detailed and systematic numerical analysis of the near-field chirality in different plasmonic nanostractures was made in Ref [6]. Importantly, the near-field chiroptical properties shown in Refs. [5, 6], are beyond the scope of the Lipkin's analysis, which was made based only on the plane wave consideration. So, an important question arises: Whether, in general, the expressions (1), (4) obtained for propagating waves are applicable for description of the chiroptical near-field response?

In a case of Eq. (4), the electric field is parallel to the magnetic field with a time-phase delay of $90°$. In Ref. [7], it was supposed that in an electromagnetic standing-wave structure, designed by interference of two counter-propagating circular polarized plane waves with the same amplitudes, there are certain planes where the electric and magnetic fields are collinear with each other and are not time-phase shifted. The authors state that such a field structure results in appearance of the energy density expressed as

$$W^{(me)} \propto \frac{1}{2} \text{Re}(\vec{E}^* \cdot \vec{B}), \quad (5)$$

which is called as the magnetoelectric (ME) energy [7]. In the vicinity of the electric nodes $kz = \pi(n+1/2)$, $n = 0,1,2,...$, one has pronounced resonances which can be interpreted as localized regions of the ME energy. Based on this analysis, authors of Ref. [7] propose a method of ultrasensitive local probing of the ME effect which can be observed in natural and artificial structures. Intuitively, it was also assumed in Ref. [7] that this ME energy density of plane monochromatic waves can be related to the reactive power flow density (or imaginary Poynting vector) [8]:

$$\vec{S}^{(me)} \propto \frac{1}{2} \text{Im}(\vec{E}^* \times \vec{B}). \quad (6)$$

For plane waves, Eq. (1) describes the local (in the sense of subwavelength scales of electromagnetic radiation) relationships between the fields and space derivatives of these fields. But the question arises for the regions of interaction with small electric and/or magnetic dipoles. In these regions, the near-field reactive energy should be taken into account and the coupling between the fields and space derivatives of the fields (in other words, the coupling between the electric and magnetic fields) appears as an effect of the first-order quantity $ka$, where $k$ is the EM wavenumber and $a$ is a particle size. Certainly, this problem of nonlocality is not overcome when we use superposition of multiple plane waves [2]. The same question of locality arises when we are talking about relations (5), (6) introduced in Ref. [7].

Recently, in Refs. [9 – 13], it was shown that the ME properties can be observed in the vacuum-region fields in the proximity of a ferrite-disk particle with MDM oscillations. Contrary to Refs. [2, 7], there are not the states of propagating-wave fields. In a case of the MDM oscillations in a small ferrite disk, we have the electric-magnetic coupling effects – the ME effects – on scales of $k_{MDM} a$, where $k_{MDM}$ is the MDM wavenumber and $a$ is a characteristic size of a ferrite particle. At the microwave frequencies, $k_{MDM} \ll k$. Free-space microwave fields, originating from magnetization dynamics in a quasi-2D ferrite disk, carry both spin and orbital angular momentums and are characterized by power-flow vortices and non-zero helicity (chirality). Topological properties of these fields – called ME fields – are different from topological properties of free-space EM fields. There are localized quantized states of the near fields [9 – 16]. A time average helicity (chirality) parameter for the near fields of the ferrite disk with MDM oscillations is defined as [9 – 13]

$$F = \frac{\varepsilon_0}{4} \text{Im} \left\{ \vec{E} \cdot \left( \vec{\nabla} \times \vec{E} \right)^* \right\} = \frac{1}{4} \omega \varepsilon_0 \mu_0 \text{Re} \left\{ \vec{E} \cdot \vec{H}^* \right\} \tag{7}$$

One can also introduce a normalized helicity parameter, which shows a time-averaged space angle between rotating vectors $\vec{E}$ and $\vec{H}$ in vacuum:

$$\cos \alpha = \frac{\text{Im} \left\{ \vec{E} \cdot \left( \vec{\nabla} \times \vec{E} \right)^* \right\}}{\left| \vec{E} \right| \left| \vec{\nabla} \times \vec{E} \right|} = \frac{\text{Re} \left\{ \vec{E} \cdot \vec{H}^* \right\}}{\left| \vec{E} \right| \left| \vec{H} \right|}. \tag{8}$$

In the near-field regions where the helicity parameter is not equal to zero, a space angle between the vectors $\vec{E}$ and $\vec{H}$ is different from π/2. Such a near-field structure breaks the field structure of Maxwell electrodynamics. The helicity states of the ME fields are topologically protected quantum-like states. It was shown [17] that at the MDM resonances, the helicity densities of the near fields are related to an imaginary part of the complex power-flow density defined as

$$\vec{S}_r = \frac{1}{2} \text{Im} \vec{E} \times \left( \vec{H} \right)^*. \tag{9}$$

The regions where the helicity density is non-zero are the regions of non-zero ME energy [17].

In studies of the ME fields, a fundamental question arises: Whether the ME-field topology observed in a subwavelength (near-field) region originated from a MDM ferrite disk can form specific far fields with unique topological characteristics? There is a very important question in a view of some topical problems of the microwave far-field radiation. As it was preliminary discussed in Ref. [7], the propagating fields with ME energy can be effectively used for sensitive probing of molecular chirality. Recently, it was shown that via microwave-radiation spectroscopy one can successfully determine the rotational energy levels of chiral molecules in the gas phase. For this study, a special experimental setup for microwave three-wave mixing was used [18]. A chiral particle is modeled as three mutually orthogonal electric-dipole moments and the handedness is identified by a sign of a triple product of these dipole moments [18]. Rotational spectroscopy is a branch of fundamental science to study the rotational spectra of molecules. Based on the rotational properties of the fields originated from MDM resonances, we can propose that novel engineered fields with ME energy can be used for remote microwave detecting and identifying the handedness of biological objects and metamaterial structures.

The near-field topological singularities originated from a MDM ferrite particle can be transmitted to the far-field region. In this paper, we report on observation of topological ME effects in the far-field microwave radiation. For study of the far-field topology we use a small microwave antenna with a MDM ferrite-disk resonator as a basic building block. At the frequency far from the MDM resonance, a ferrite disk appears as a small obstacle in a waveguide and our microwave structure behaves as usual waveguide-hall antenna. The situation is cardinally changed when we are at the MDM resonance frequency. In such a case, we observe the helicity density in the far-field structure. The helicity densities of the far fields are related to an imaginary part of the complex power-flow density. The observed effects of far-field microwave transportation of ME energy allow better understanding of interaction between MDM magnons and microwave radiation.

## II. LOCALIZED REGIONS OF THE ME ENERGY: THE ME FIELDS NEAR A FERRITE DISK WITH MDM SPECTRA

Following an analysis in Ref. [7] on the localized regions of the ME fields, let us suppose that the fields $\vec{E}^{(1)}, \vec{H}^{(1)}$ and $\vec{E}^{(2)}, \vec{H}^{(2)}$ are two counterpropagating plane waves with the same amplitudes and circular polarizations of the same direction:

$$\vec{E}^{(1)} = E_0(\hat{\vec{x}} + i\sigma\hat{\vec{y}})e^{-ikz}, \quad \vec{H}^{(1)} = H_0(\hat{\vec{y}} - i\sigma\hat{\vec{x}})e^{-ikz},$$

$$\vec{E}^{(2)} = E_0(\hat{\vec{x}} + i\sigma\hat{\vec{y}})e^{ikz}, \quad H^{(2)} = -H_0(\hat{\vec{y}} - i\sigma\hat{\vec{x}})e^{ikz}, \qquad (11)$$

where $\sigma = \pm 1$ determines the circular-polarization direction. For a standing wave formed by these two fields we have

$$\vec{E} = \vec{E}^{(1)} + \vec{E}^{(2)} = E_0\left[(\hat{\vec{x}} + i\sigma\hat{\vec{y}})e^{-ikz} + (\hat{\vec{x}} + i\sigma\hat{\vec{y}})e^{ikz}\right],$$

$$\vec{H} = \vec{H}^{(1)} + \vec{H}^{(2)} = H_0\left[(\hat{\vec{y}} - i\sigma\hat{\vec{x}})e^{-ikz} - (\hat{\vec{y}} - i\sigma\hat{\vec{x}})e^{ikz}\right]. \qquad (12)$$

From such a field configuration, we obtain

$$\frac{1}{2}\text{Re}\left(\vec{E}\cdot\vec{H}^*\right) = -2\sigma E_0 H_0 \sin(2kz). \qquad (13)$$

In the vicinity of the electric nodes $kz = \pi(n+1/2)$, $n = 0,1,2,...$, one has pronounced resonances which can be interpreted as localized regions of the ME energy. Based on this analysis, authors of Ref. [7] propose a method of ultrasensitive local probing of the ME effect which can be observed in natural and artificial structures.

The above statement that based on only two EM plane waves in vacuum counterpropagating with the same circular polarizations one can obtain localized regions of the ME energy raises, however, the questions of a fundamental character. This assertion appears more as a mathematical, but not a physically realizable concept. In observation of the ME phenomena by pure EM means, the problem of nonlocality plays the underlying role. In other words, for pure EM interactions, "magnetoelictricity" appears only due to field nonlocality. Optically active media, chiral biological materials, and bianisotropic metamaterials are spatially dispersive structures. The ME properties are exhibited as the effect of the first-order quantity *ka*, where *k* is the EM wavenumber and *a* is a characteristic size (the particle size or the distance between particles) [19 – 21]. Evidently, this problem of nonlocality cannot be overcome when one uses superposition of multiple plane waves [7].

The ME energy, defined by Eq. (13), transforms as a pseudo-scalar under space reflection $\mathcal{P}$ and it is odd under time reversal $\mathcal{T}$. When such a quadratic form exists, the Maxwell electrodynamics should be extended to so-called axion-electrodynamics [22]. Axion electrodynamics, i.e., the standard electrodynamics modified by an additional axion field, provides a theoretical framework for a possible violation of parity and Lorentz invariance. An axion-electrodynamics term, added to the ordinary Maxwell Lagrangian [22]:

$$\mathcal{L}_\vartheta = \kappa \vartheta \vec{E} \cdot \vec{B}, \qquad (14)$$

where $\kappa$ is a coupling constant, results in modified electrodynamics equations with the electric charge and current densities replaced by [22, 23]

$$\rho^{(e)} \to \rho^{(e)} + \kappa \vec{\nabla} \vartheta \cdot \vec{B}, \qquad (15)$$

$$\vec{j}^{(e)} \to \vec{j}^{(e)} - \kappa \left( \frac{\partial \vartheta}{\partial t} \vec{B} + \vec{\nabla} \vartheta \times \vec{E} \right). \qquad (16)$$

The form of these terms reflects the discrete symmetries of $\vartheta$: $\vartheta$ is $\mathcal{P}$ and $\mathcal{T}$ odd. A dynamical pseudoscalar field $\vartheta$, called an axion field, which couples to $\vec{E} \cdot \vec{B}$. The $\vartheta$ term is topological in the sense that it does not depend on the space-time metric. Whenever a pseudo-scalar axion-like field $\vartheta$ is introduced in the theory, the dual symmetry between the electric and magnetic fields is spontaneously and explicitly broken. The form of the effective action implies that an electric field can induce a magnetic polarization, whereas a magnetic field can induce an electric polarization. This effect is known as the topological ME effect and $\vartheta$ is related to the ME polarization. Axion-like fields and their interaction with the electromagnetic fields have been intensively studied and they have recently received attention due to their possible role played in building topological insulators [24 – 28]. A topological isolator is characterized by additional parameter $\vartheta$ that couples the pseudo-scalar product of the electric and magnetic fields in the effective topological action. For $\vartheta = \pi$, the axion-electrodynamics Lagrangian describes the unusual ME properties of the topological isolator. In some materials, the axion field couples linearly to light, resulting in the axionic polariton. Contrary to optically active media, chiral biological materials, and bianisotropic metamaterials, the topological ME is basically a surface effect rather than a bulk one. Spin 1/2 particles exhibit the counterintuitive property that their wave function acquires the $\pi$ phase upon $2\pi$ rotation. If spin and orbital degrees of freedom are mixed in a particular way, the momenta can feel important effects of this $\pi$ Berry's phase, which can lead to a new phase of waves whose description requires a fundamental redress of the theory of electromagnetic radiation.

In contrast to a formal representation of the "ME energy density" in Ref. [7], expressed by Eqs. (5), (13), the helicity parameter $F$ for the near fields of a ferrite particle with MDM oscillations [described by Eq. (7)] is justified physically as a quantity related to an axion-electrodynamics term [9, 10, 13 – 15]. A ferrite material, by itself is not a ME material. In a case of a MDM ferrite disk, the ME coupling is the topological effect which arises through the chiral edge states. For MDMs, the magnetic field is a potential field: $\vec{H} = -\vec{\nabla}\psi$, where $\psi$ is the magnetostatic-potential (MS-potential) wave function. In an assumption of separation of variables, a magnetostatic-potential (MS-potential) wave function in a ferrite disk is represented in cylindrical coordinates $z, r, \theta$ as $\psi(r,\theta,z) = C\xi(z)\tilde{\varphi}(r,\theta)$, where $\tilde{\varphi}$ is a dimensionless membrane MS-potential wave function, $\xi(z)$ is a dimensionless amplitude factor, and $C$ is a dimensional coefficient. On a lateral surface of a quasi-2D ferrite disk of radius $\Re$, a MS-potential membrane wave function is expressed as: $\left(\tilde{\varphi}_{\pm}\right)_{r=\Re^-} = \delta_{\pm}\left(\tilde{\eta}\right)_{r=\Re^-}$, $\tilde{\eta}$ is a singlevalued membrane function and $\delta_{\pm}$ is a double-valued edge wave function on contour $\mathcal{L} = 2\pi\Re$. Function $\delta_{\pm}$ changes its sign when the regular-coordinate angle $\theta$ is rotated by $2\pi$. As a result, one has the eigenstate spectrum of MDM oscillations with topological phases accumulated by the edge wave function $\delta$. On a lateral surface of a quasi-2D ferrite disk, one can distinguish two different functions $\delta_{\pm}$, which are the counterclockwise and clockwise rotating-wave edge functions with respect to a membrane function $\tilde{\eta}$. A line integral around a singular contour $\mathcal{L}$:

$\frac{1}{\Re}\oint_{\mathcal{L}}\left(i\frac{\partial\delta_{\pm}}{\partial\theta}\right)(\delta_{\pm})^* d\mathcal{L} = \int_0^{2\pi}\left[\left(i\frac{\partial\delta_{\pm}}{\partial\theta}\right)(\delta_{\pm})^*\right]_{r=\Re} d\theta$ is an observable quantity. Because of the existing the geometrical phase factor on a lateral boundary of a ferrite disk, MDMs are characterized by a pseudo-electric field (the gauge field) $\vec{\epsilon}$. The pseudo-electric field $\vec{\epsilon}$ can be found as $\vec{\epsilon}_{\pm} = -\vec{\nabla}\times\left(\vec{\Lambda}_{\epsilon}^{(m)}\right)_{\pm}$. The field $\vec{\epsilon}$ is the Berry curvature. The corresponding flux of the gauge field $\vec{\epsilon}$ through a circle of radius $\Re$ is obtained as: $K\int_S\left(\vec{\epsilon}\right)_{\pm}\cdot d\vec{S} = K\oint_{\mathcal{L}}\left(\vec{\Lambda}_{\epsilon}^{(m)}\right)_{\pm}\cdot d\vec{\mathcal{L}} = K\left(\Xi^{(e)}\right)_{\pm} = 2\pi q_{\pm}$, where $\left(\Xi^{(e)}\right)_{\pm}$ are quantized fluxes of pseudo-electric fields. Each MDM is quantized to a quantum of an emergent electric flux. There are the positive and negative eigenfluxes. These different-sign fluxes should be nonequivalent to avoid the cancellation. It is evident that while integration of the Berry curvature over the regular-coordinate angle $\theta$ is quantized in units of $2\pi$, integration over the spin-coordinate angle $\theta'$ $\left(\theta' = \frac{1}{2}\theta\right)$ is quantized in units of $\pi$. The physical meaning of coefficient $K$ concerns the property of a flux of a pseudo-electric field. The Berry mechanism provides a microscopic basis for the surface magnetic current at the interface between gyrotropic and nongyrotropic media. Following the spectrum analysis of MDMs in a quasi-2D ferrite disk one obtains pseudo-scalar axion-like fields and edge chiral magnetic currents. Topological properties of ME fields (non-zero helicity factor) arise from the presence of geometric phases on a border circle of a MDM ferrite disk.

In the near-field vacuum area of a quasi-2D ferrite disk with MDM resonances, one has in-plane rotating electric- and magnetic-field vectors localized at a center of a disk [21]. This field structure, shown schematically in Fig. 1, is characterized by the helicity factor, which is related to the product $\vec{E} \cdot \vec{B}$. In a numerical analysis we use the yttrium iron garnet (YIG) disk of diameter of 3 mm. The disk thickness is 0.05 mm. The disk is normally magnetized by a bias magnetic field $H_0 = 4760$ Oe; the saturation magnetization of a ferrite is $4\pi M_s = 1880$ G. For better understanding the field structures, in a numerical analysis we use a ferrite disk with a very small linewidth of $\Delta H = 0.1$ Oe. The numerically calculated helicity density for the main MDM is shown in Fig.1 for two opposite directions of a bias magnetic field.

While, for MDMs, the magnetic field in a vacuum region near a ferrite disk is a potential field, $\vec{H} = -\vec{\nabla}\psi$, the electric field has two parts: the curl-field component $\vec{E}_c$ and the potential-field component $\vec{E}_p$. The curl electric field $\vec{E}_c$ in vacuum we define from the Maxwell equation $\vec{\nabla} \times \vec{E}_c = -\mu_0 \frac{\partial \vec{H}}{\partial t}$. The potential electric field $\vec{E}_p$ in vacuum is calculated by integration over the ferrite-disk region, where the sources (magnetic currents $\vec{j}^{(m)} = \frac{\partial \vec{m}}{\partial t}$) are given. Here $\vec{m}$ is dynamical magnetization in a ferrite disk. Eq. (7) for the helicity density of the ME field is non-zero when one has non-zero scalar product $\vec{E}_p \cdot (\vec{\nabla} \times \vec{E}_c)$ [9, 10, 13 – 15]. With representation of the potential electric field as $\vec{E}_p = -\vec{\nabla}\phi$, we can write for ME fields in vacuum:

$$F = \frac{\omega_{MDM}\varepsilon_0}{4}\text{Re}\left(\vec{E} \cdot \vec{B}^*\right) = -\frac{\omega_{MDM}\varepsilon_0\mu_0}{4}\text{Re}\left[\vec{\nabla}\phi \cdot \left(\vec{\nabla}\psi\right)^*\right]. \tag{17}$$

In Ref. [17], it was shown that the vacuum regions above and below a quasi-2D ferrite disk, where the helicity density parameter $F$ is nonzero, are also the regions where the imaginary part of a vector $\vec{E} \times \vec{H}^*$ exist as well. For the near-field vacuum areas, localized at an axis of a quasi-2D ferrite disk, this connection is represented by the following relation:

$$\frac{1}{2}\text{Re}\left|\vec{E} \cdot \vec{H}^*\right| = \frac{1}{2}\text{Im}\left|\left[\vec{E} \times \left(\vec{H}\right)^*\right]_z\right|. \tag{18}$$

The right-hand side of this equation describes a projection of the reactive power flow on the axis of a ferrite disk. Fig. 2 shows ME-field power flows near a ferrite disk for the main MDM oscillation.

When a MDM ferrite disk is placed inside a microwave structure, a time-averaged space angle between rotating vectors $\vec{E}$ and $\vec{H}$ in vacuum [described by Eq. (8)] varies because of a role of the curl fields of a microwave structure. This variation of an angle between spinning electric and magnetic fields along the disk $z$-axis for the main MDM is shown in Fig. 3. This angle gives evidence for a torsion structure of the ME field above and below a ferrite disk. The ME-energy density appears due to the torsion degree of freedom of the field.

**III. FAR-FIELD TOPOLOGICAL ME EFFECTS**

Axion-like fields can interact with the EM fields, but cannot be eliminated by the EM fields. It means that the torsion degree of freedom observed due to an axion-electrodynamics term in the near-field region of a MDM ferrite resonator, cannot be removed "electromagnetically" in the far-field region of microwave radiation. To find the transfer of the topological ME effects in the far-field region, a microwave antenna should not have other resonant elements except the MDM resonant structure. For this purpose, we use a rectangular waveguide with a hole in a wide wall and the diameter of this hole is much less than a half wavelength of microwave radiation. At the MDM resonance, the topological singularities in the radiation field appear due to topological singularities in the electric current distributions on the external surface of a waveguide wall. These current singularities are well distinguished in a microwave antenna with symmetrical geometry. For this reason, the hole in a waveguide wide wall is situated symmetrically.

A microwave antenna used in our studies is shown in Fig. 4. This is a waveguide radiation structure with a hole in a wide wall and a thin-film ferrite disk as a basic building block. A ferrite disk is placed inside a $TE_{10}$-mode rectangular X-band waveguide symmetrically to its walls so that a disk axis is perpendicular to a wide wall of a waveguide. For numerical studies, the disk has the same parameters as discussed in Section II. The waveguide walls are made of a perfect electric conductor (PEC). A hole in a wide wall has a diameter of 8 mm, which is much less than a half wavelength of microwave radiation at the frequency regions from 8.0 GHz till 8.5 GHz, used in our studies.

Fig. 5 shows the numerical reflection characteristics for the $TE_{10}$ waveguide mode in our microwave antenna. The spectrum is quite emasculated compared to a multiresonant spectrum of MDM oscillations observed in non-radiating microwave structures [9 – 12]. Because of the presence of a radiation hole, high-order MDMs are not excited quite effectively. Numerically, we primarily observe excitation of the first (main) MDM oscillation. The forms of the resonance peaks in Fig. 5 give evidence for the Fano-type interaction. The underlying physics of the Fano resonances finds its origin in wave interference which occurs in the systems characterized by discrete energy states that interact with the continuum spectrum (an entire continuum composed by the internal waveguide and external free-space regions). The forward and backward propagating modes within the waveguide are coupled via the defects. This coupling becomes highly sensitive to the resonant properties of the defect states. For such a case, the coupling can be associated with the Fano resonances. In the corresponding transmission dependencies, the interference effect leads to either perfect transmission or perfect reflection, producing a sharp asymmetric response. We have a "bright" and a "dark" resonances, which produce the Fano-resonance form in the reflection spectra. In our study we will use the first "bright" peak – the peak $1'$ – where the most intensive radiation is observed. At the resonance peak $1'$ frequency the field structure near a ferrite disk is typical for the main MDM excited in a closed waveguide system [9 – 12, 17]: there are rotating electric and magnetic field, the active power flow vortex, the reactive power flow and the reactive power flow. Far-field topological properties of ME fields arise from the presence of geometric phases in the radiation near-field region. These geometric phases should have opposite signs for the frequencies situated to the left and to the right from the resonance frequency of peak $1'$. The resonance frequency of the peak $1'$ in Fig. 5 is $f$ = 8.139GHz. To show a role of geometric phases in the far-field radiation, in the further analysis we will use the frequencies $f$ = 8.138GHz and $f$ = 8.140GHz. There are, respectively, the frequencies on the left and right slopes of the resonance peak, very close to the top of this resonance peak.

To observe the far-field topological ME effects originated from a MDM ferrite particle, we should create the regions with localized intensities of the electric and magnetic fields in vacuum. For this purpose, we place a dielectric ($\varepsilon_r = 10$) plate above the antenna. The plate with sizes $40 \times 40 \times 1$ mm is disposed parallel to the waveguide surface at distance about $3.5\lambda_0$, where $\lambda_0$ is the electromagnetic wavelength in vacuum. Fig. 6 shows a normalized helicity parameter in a vacuum far-field region calculated based on Eq. (8). At the frequency $f$ = 8.138GHz one observes striations of the positive and negative quantities of the normalized helicity parameter. Periodicity of the striations is associated with the standing-wave behavior of regular EM waves in vacuum.

Since a sign of a dynamical pseudoscalar field $\vartheta$ in Eq. (14) is correlated with a direction of a bias magnetic field applied to a ferrite disk, the helicity-parameter striations change their signs, when a bias magnetic field is in an opposite direction.

Fig. 7 represents a detailed analysis of the helicity parameter distribution in a vacuum far-field region between an antenna and a dielectric plate. Figs. 7 (*a*), (*b*), show the regions of localization of the helicity parameter in different vacuum horizontal planes. From Figs. 7 (c) (d), we can see how the helicity parameter is correlated with the regions of localized reactive power flows. As it was shown in Ref. [17] at the MDM resonances, the regions of localization of ME energy are correlated with the regions of the localized field intensities and the regions of localized reactive power flows near a ferrite disk. The similar situation we have in a far-field region. Along *z*-axis, the maximums of the helicity-parameter modulus are at the same positions as the regions of maximal gradient modulus of the reactive-power flow. Also, we can see that the maximums of the helicity-parameter modulus are at the regions of maximal gradient modulus of the field (both electric and magnetic) intensities. The positions of striations with the positive and negative helicity parameters are correlated with the EM standing-wave structure. For a given radiation frequency (*f* = 8.138GHz), these positions remain the same when one uses a dielectric plate with another permittivity parameters, but the same sizes. When the intensity of the electric and magnetic fields in the EM standing-wave structure increases the intensity of the striations increases as well. Fig. 8 shows the pictures of the helicity parameter distributions for two different permittivity parameters of the dielectric plates.

Striations of the positive and negative quantities of the normalized helicity parameter are regions of the positive and negative ME energy. In these regions, the electric and magnetic fields have components that are mutually parallel (antiparallel) and are not time-phase shifted. At the same time, small chiral particles being excited by an electromagnetic field have the electric field parallel to the magnetic field but with a time-phase delay of 90° [2]. Since the ME field originated from a MDM ferrite disk and the field originated from a chiral particle are fundamentally different, no strong perturbation of the ME-field structure by chiral particles should be observed. Fig. 9 illustrates this situation. In this figure, an array of small metallic helices, oriented along *z* axis, is placed in space between an antenna and a dielectric plate ($\varepsilon_r = 10$). The helix diameter is 1.5mm and the height is 2.0mm. The helices are made of a PEC wire with diameter of 0.1mm. The number of helix turns is five. One can see that the helices do not change positions of the helicity-parameter striations. However, the regions inside every helix become green colored. It means that the electric and magnetic fields inside helices become time-phase shifted with 90°. So, inside helices we have restoration of a regular EM-field behavior.

As we discussed above, topological properties of ME fields arise from the presence of geometric phases in MDM oscillations. If it is so, the change of a sign of a dynamical pseudoscalar field in Eq. (14) will simultaneously change a sign of a scalar product $\vec{E} \cdot \vec{B}$. All the above results were obtained for the frequency *f* = 8.138GHz, which on the left slope of the resonance peak. When we go to the frequency *f* = 8.140GHz, situated on right slope of the resonance peak, we will have an opposite sign of a geometric phase leading to change a sign of a helicity factor – a sign of a scalar product $\vec{E} \cdot \vec{B}$. Fig. 10 shows far-field helicity-factor distributions for the frequency *f* = 8.138GHz (the left slope of the resonance peak) and the frequency *f* = 8.140GHz (the right slope of the resonance peak). The selected regions show the helicity-factor distributions near a ferrite disk and near an antenna aperture. Near a ferrite disk, the ME field is very strong compared to the EM field in a waveguide. In this region, no ME-EM field interaction takes place and a sign of a helicity factor is determined only by a direction of a bias magnetic field. However, near an antenna aperture one observes the ME-EM field interaction. In this near-field region, the helicity-factor distributions are different for the frequencies *f* = 8.138GHz and *f* = 8.140GHz. Different helicity-factor distributions in the radiation near-field region are origins of different helicity-factor distributions in the far-field region.

For the optical chirality $\chi$, expressed by Eqs. (1), it is evident that for monochromatic electromagnetic waves we have

$$\mathrm{Im}\,\chi = \mathrm{Im}\left[\frac{\varepsilon_0}{4}\vec{E}\cdot\left(\nabla\times\vec{E}\right)^* + \frac{\mu_0}{4}\vec{H}\cdot\left(\nabla\times\vec{H}\right)^*\right] = \frac{\omega\varepsilon_0\mu_0}{2}\mathrm{Re}\left(\vec{E}^*\cdot\vec{H}\right) \equiv 0. \qquad (19)$$

Let us denote, formally, the terms in this equation as follows

$$F^{(E)} = \mathrm{Im}\left[\frac{\varepsilon_0}{4}\vec{E}\cdot\left(\nabla\times\vec{E}\right)^*\right] \quad \text{and} \quad F^{(H)} = \mathrm{Im}\left[\frac{\mu_0}{4}\vec{H}\cdot\left(\nabla\times\vec{H}\right)^*\right]. \qquad (20)$$

For monochromatic EM waves, we have, evidently

$$F^{(E)} \equiv 0 \text{ and } F^{(H)} \equiv 0. \qquad (21)$$

At the same time, for the ME near fields, we have [9]

$$F^{(E)} \equiv F = \frac{\varepsilon_0}{4}\mathrm{Im}\left\{\vec{E}\cdot\left(\vec{\nabla}\times\vec{E}\right)^*\right\} = \frac{1}{4}\omega\varepsilon_0\mu_0\,\mathrm{Re}\left\{\vec{E}\cdot\vec{H}^*\right\} \neq 0 \quad \text{and} \quad F^{(H)} \equiv 0. \qquad (22)$$

If we suppose that for the ME far fields both $F^{(E)} \neq 0$ and $F^{(H)} \neq 0$, we will have

$$F^{(E)} = \frac{\varepsilon_0}{4}\mathrm{Im}\left[\vec{E}\cdot\left(\nabla\times\vec{E}\right)^*\right] = \frac{\omega\varepsilon_0\mu_0}{4}\mathrm{Re}\left(\vec{E}\cdot\vec{H}^*\right), \qquad (23)$$

$$F^{(H)} = \frac{\mu_0}{4}\mathrm{Im}\left[\vec{H}\cdot\left(\nabla\times\vec{H}\right)^*\right] = -\frac{\omega\varepsilon_0\mu_0}{4}\mathrm{Re}\left(\vec{H}\cdot\vec{E}^*\right) = -\frac{\omega\varepsilon_0\mu_0}{4}\mathrm{Re}\left(\vec{E}\cdot\vec{H}^*\right). \qquad (24)$$

So,

$$F^{(E)} = -F^{(H)}. \qquad (25)$$

This gives

$$F^{(E)} + F^{(H)} = 0. \qquad (26)$$

The above formal consideration can be verified by the following analysis of transfer of the topological ME effects in the far-field region. As we can see from Eq. (17), in the near-field region both the electric and magnetic fields are potential fields. There are the topological-nature fields. If we assume that such potential (topological) fields also exist in the far-field radiation, we should consider the total fields as the composite fields containing both the curl and potential (topological) field components:

$$\vec{E}_t = \vec{E}_c + \vec{E}_p \quad \text{and} \quad \vec{H}_t = \vec{H}_c + \vec{H}_p. \qquad (27)$$

For these fields components we have the equations:

$$\vec{\nabla} \times \vec{E}_c = -\mu_0 \frac{\partial \vec{H}_t}{\partial t}, \qquad \vec{\nabla} \times \vec{E}_p = 0,$$

$$\vec{\nabla} \times \vec{H}_c = \varepsilon_0 \frac{\partial \vec{E}_t}{\partial t}, \qquad \vec{\nabla} \times \vec{H}_p = 0. \qquad (28)$$

The curl fields, $\vec{E}_c$ and $\vec{H}_c$ are the Maxwellian fields and the potential fields, $\vec{E}_p$ and $\vec{H}_p$ are topological fields originated from a MDM particle. For time harmonic fields in the far-field region we can write:

$$F^{(E)} = \frac{\varepsilon_0}{2} \left[ \left( \vec{E}_c + \vec{E}_p \right) \cdot \nabla \times \vec{E}_c \right] = -i \frac{\omega \varepsilon_0 \mu_0}{2} \left( \vec{E}_c + \vec{E}_p \right) \cdot \left( \vec{H}_c + \vec{H}_p \right), \qquad (29)$$

$$F^{(H)} = \frac{\mu_0}{2} \left[ \left( \vec{H}_c + \vec{H}_p \right) \cdot \nabla \times \vec{H}_c \right] = i \frac{\omega \varepsilon_0 \mu_0}{2} \left( \vec{H}_c + \vec{H}_p \right) \cdot \left( \vec{E}_c + \vec{E}_p \right). \qquad (30)$$

Taking into account that $\vec{E}_c \cdot \vec{H}_c \equiv 0$, we obtain

$$F^{(E)} = \frac{\varepsilon_0}{4} \operatorname{Im} \left[ \vec{E}_p \cdot \left( \nabla \times \vec{E}_c \right)^* \right] = \frac{\omega \varepsilon_0 \mu_0}{4} \operatorname{Re} \left( \vec{E}_p \cdot \vec{H}_c^* + \vec{E}_p \cdot \vec{H}_p^* \right), \qquad (31)$$

$$F^{(H)} = \frac{\mu_0}{4} \left[ \vec{H}_p \cdot \left( \nabla \times \vec{H}_c \right)^* \right] = -\frac{\omega \varepsilon_0 \mu_0}{4} \operatorname{Re} \left( \vec{H}_p \cdot \vec{E}_c^* + \vec{H}_p \cdot \vec{E}_p^* \right)$$
$$= -\frac{\omega \varepsilon_0 \mu_0}{4} \operatorname{Re} \left( \vec{E}_c \cdot \vec{H}_p^* + \vec{E}_p \cdot \vec{H}_p^* \right). \qquad (32)$$

Assuming Eq. (26), we have

$$\operatorname{Re} \left( \vec{E}_p \cdot \vec{H}_c^* \right) = -\operatorname{Re} \left( \vec{E}_c \cdot \vec{H}_p^* \right). \qquad (33)$$

This analysis of the topological ME effect shows that in the far-field region we have two kinds of the torsion fields. There the "electric-torsion" fields characterizing by the helicity parameter $F^{(E)}$ and the "magnetic-torsion" fields characterizing by the helicity parameter $F^{(H)}$. For a given sign of a dynamical pseudoscalar field $\vartheta$, an axion-electrodynamics term, expressed by Eq. (14), has different signs for the "electric" and "magnetic" helicities. Fig. 11 shows normalized helicity parameters in a vacuum far-field region at the frequency $f = 8.138 \text{GHz}$, calculated based on the equations:

$$\cos \alpha = \frac{\operatorname{Im} \left\{ \vec{E} \cdot \left( \vec{\nabla} \times \vec{E} \right)^* \right\}}{\left| \vec{E} \right| \left| \vec{\nabla} \times \vec{E} \right|} \quad \text{and} \quad \cos \beta = \frac{\operatorname{Im} \left\{ \vec{H} \cdot \left( \vec{\nabla} \times \vec{H} \right)^* \right\}}{\left| \vec{H} \right| \left| \vec{\nabla} \times \vec{H} \right|}. \qquad (34)$$

In spite of the fact that both the "electric" and "magnetic" helicities exist, they annihilate one another. Fig. 12 illustrates this situation. In the far-field region, the turn of a total electric field $\vec{E}_t = \vec{E}_c + \vec{E}_p$ with respect to a curl magnetic field $\vec{H}_c$ is compensated by the turn of a total magnetic field $\vec{H}_t = \vec{H}_c + \vec{H}_p$ with respect to a curl electric field $\vec{E}_c$. This results in mutual perpendicularity between $\vec{E}_t$ and $\vec{H}_t$. In a local point in the far-field region, the "red" and "blue" helicity factors being superposed one to another, give the "green" (zero) helicity factor. In other words, the negative ME energy is compensated by the positive ME energy. These local properties of the "helicity neutrality" mean that in the far-field region the field structure, being composed with the "electric-torsion" and "electric-torsion" components becomes non-distinguishable from the regular EM-field structure.

The standard means do not give possibility to observe the ME-field structure in the far-field region. The only possibility is to use another MDM ferrite disk as an indicator. This ferrite disk will, definitely, distinguish the "electric" helicity from the "magnetic" helicity.

## IV. CONCLUSION

In this paper we analyzed interaction between MDM magnons and far-field microwave radiation. We showed that the near-field topological singularities originated from a MDM ferrite particle can be transmitted to the far-field region. For study of the far-field topology we used a small microwave antenna with a MDM ferrite-disk resonator as a basic building block. At the frequency far from the MDM resonance, a ferrite disk appears as a small obstacle in a waveguide and our microwave structure behaves as usual waveguide-hall antenna. The situation is cardinally changed when we are at the MDM resonance frequency.

The microwave far-field radiation can be manifested with a torsion structure where an angle between the electric and magnetic field vectors varies. We discussed the question on observation of the regions of localized ME energy in far-field microwave radiation. The standard means do not give possibility to observe the ME-field structure in the far-field region. This our statement contradicts to the assertion in Ref. [7], that one can observe the ME-field regions for propagating EM plain waves. We conclude that the only way to observe the ME-field structure in the far-field region is to use another MDM ferrite disk as an indicator. In Ref. [11] we showed that interactions between two MDM disks are manifested at distances much larger than the EM wavelength. The frequency region of these interactions is within the scales of the MDM resonance peak width. Our future studies are aimed to use a MDM ferrite antenna together with a MDM indicator to show experimentally the ME-field structure in the far-field region of microwave radiation.

**References**
[1] D. M. Lipkin, J. Math. Phys. **5**, 696 (1964).
[2] Y. Tang and A. E. Cohen, Phys. Rev. Lett. **104**, 163901 (2010).
[3] K. Bliokh and F. Nori, Phys. Rev. A **83**, 021803(R) (2011).
[4] J. S. Choi and M. Cho, Phys. Rev. A **86**, 063834 (2012).
[5] E. Hendry *et al*, Nat. Nanotechnol. **5**, 783 (2010).
[6] M. Schäferling, D. Dregely, M. Hentschel, and H. Giessen, Phys. Rev. X **2**, 031010 (2012).
[7] K. Y. Bliokh, Y. S. Kivshar and F. Nori, Phys. Rev. Lett. **113**, 033601 (2014).
[8] J. D. Jackson, *Classical Electrodynamics*, 2nd ed. (Wiley, New York, 1975).
[9] E.O. Kamenetskii, R. Joffe, R. Shavit, Phys. Rev. E **87**, 023201 (2013).
[10] R. Joffe, E.O. Kamenetskii and R. Shavit, J. Appl. Phys. **113**, 063912 (2013).
[11] M. Berezin, E. O. Kamenetskii, and R. Shavit, J. Opt. **14**, 125602 (2012).
[12] M. Berezin, E. O. Kamenetskii, and R. Shavit, Phys. Rev. E **89**, 023207 (2014).

[13] E. O. Kamenetskii, E. Hollander, R. Joffe, R. Shavit, J. Opt. **17**, 025601 (2015).
[14] E. O. Kamenetskii, J. Phys. A: Math. Theor. **40**, 6539 (2007).
[15] E. O. Kamenetskii, J. Phys.: Condens. Matter **22**, 486005 (2010).
[16] E. O. Kamenetskii, R. Joffe, and R. Shavit, Phys. Rev. A **84**, 023836 (2011).
[17] E. O. Kamenetskii, M. Berezin, R. Shavit, arXiv:1502.00220.
[18] D. Patterson, M. Schnell, and J. M. Doyle, Nature **497**, 475 (2013).
[19] L.D. Landau and E.M. Lifshitz, *Electrodynamics of Continuous Media*, 2nd ed. (Pergamon, Oxford, 1984).
[20] V.M. Agranovich and V.L. Ginzburg, *Crystal Optics with Spatial Dispersion and Excitons*, 2nd ed. (Springer-Verlag, New York, 1984).
[21] E.O. Kamenetskii, M. Sigalov, and R. Shavit, J. Appl. Phys. **105**, 013537 (2009).
[22] F. Wilczek, Phys. Rev. Lett. **58**, 1799 (1987).
[23] L. Visinelli, arXiv:1111.2268 (2011).
[24] X.-L. Qi, T. L. Hughes, and S.-C. Zhang, Phys. Rev. B **78**, 195424 (2008).
[25] A. M. Essin, J. E. Moore, and D. Vanderbilt, Phys. Rev. Lett. **102**, 146805 (2009).
[26] M. Z. Hasan and C. L. Kane, Rev. Mod. Phys. **82**, 3045 (2010).
[27] R. Li, J. Wang, X.-L. Qi, and S.-C. Zhang, Nature Phys. **6**, 284 (2010).
[28] H. Ooguri and M. Oshikawa, Phys. Rev. Lett. **108**, 161803 (2012).

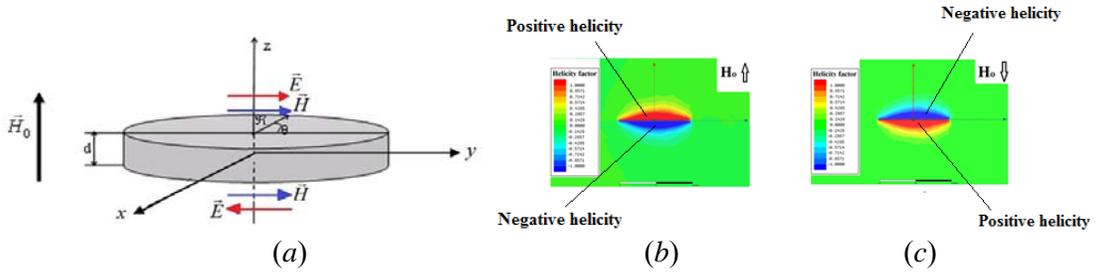

Fig. 1. The topological eigen characteristics of a ferrite disk for the main MDM oscillation. (*a*) Spinning electric- and magnetic-field vectors in vacuum regions above and below a MDM-resonance ferrite disk. (*b*) and (*c*) the helicity factor. Evidently, we have a topological "helicity dipole", which is aligned with the bias magnetic field. It means that ME properties are non-degenerate with respect to the direction of the magnetic field. So, the MDM ferrite disk, being placed in a normal DC magnetic field, has a ME energy. Also, a topological "helicity dipole" is spatially anti-symmetric with respect to the disk middle plane. The MDM ferrite disk behaves as a $\mathcal{PT}$-symmetrical structure.

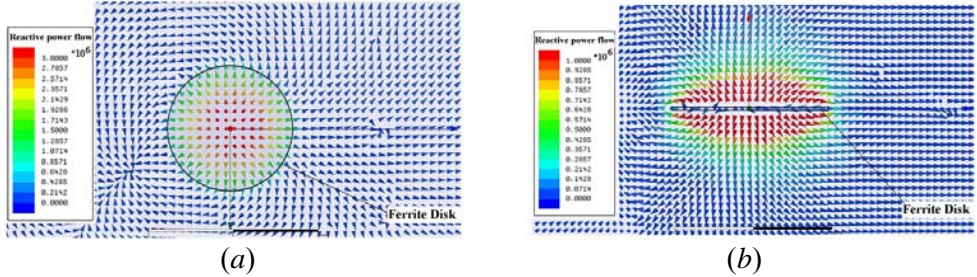

Fig. 2. ME-field reactive power flow near a ferrite disk for the main MDM oscillation. (*a*) A view on a vacuum plane parallel to the ferrite-disk plane above a disk, (*b*) a view on a cross-section plane perpendicular to the ferrite-disk plane.

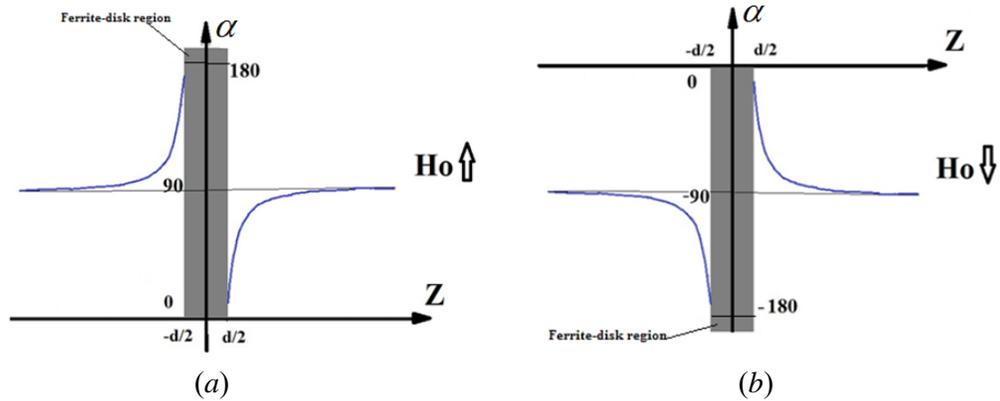

Fig. 3. Variation of the angle between spinning electric and magnetic fields along the disk *z*-axis for the main MDM. This angle gives evidence for a torsion structure of the ME field above and below a ferrite disk. The ME-energy density appears due to the torsion degree of freedom of the field. (*a*) Angle $\alpha$ for an upward directed bias magnetic field; (*b*) angle $\alpha$ for a downward directed bias magnetic field.

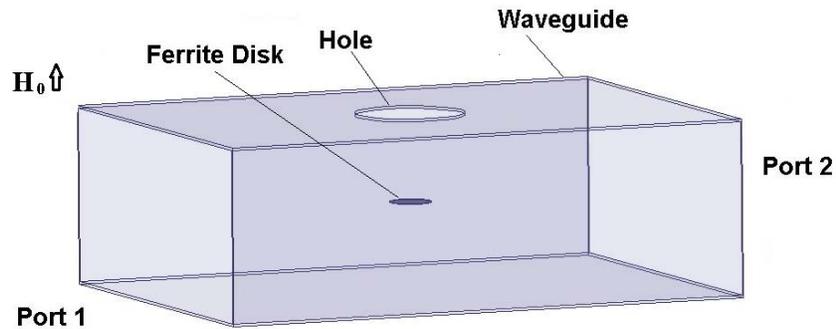

Fig. 4. MDM microwave antenna: a waveguide radiation structure with a hole in a wide wall and a thin-film ferrite disk as a basic building block. A ferrite disk is placed inside a $TE_{10}$-mode rectangular X-band waveguide symmetrically to its walls so that a disk axis is perpendicular to a wide wall of a waveguide. A hole diameter is much less than a half wavelength of microwave radiation at the frequency regions used in the studies.

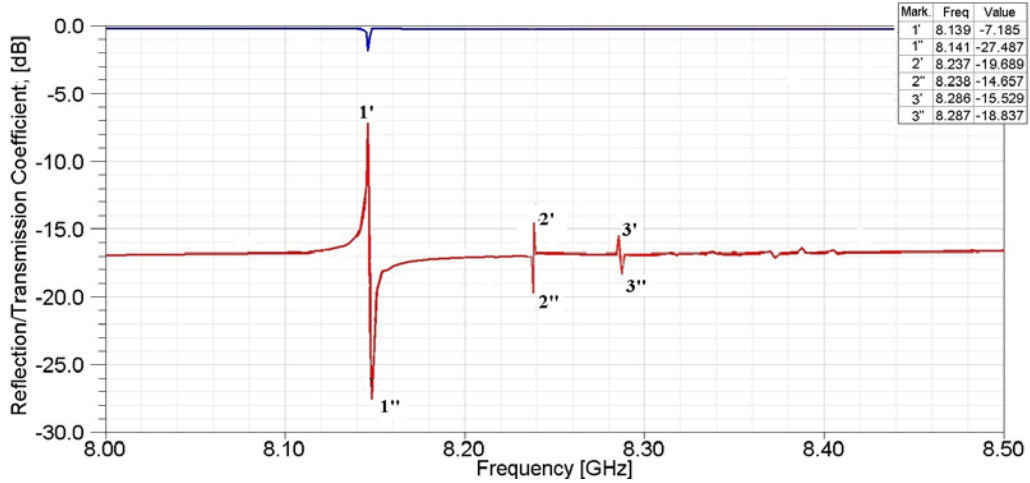

Fig. 5. Numerical reflection characteristic. Because of the presence of a radiation hole, high-order MDMs are not excited quite effectively. The forms of the resonance peaks give evidence for the Fano-type interaction.

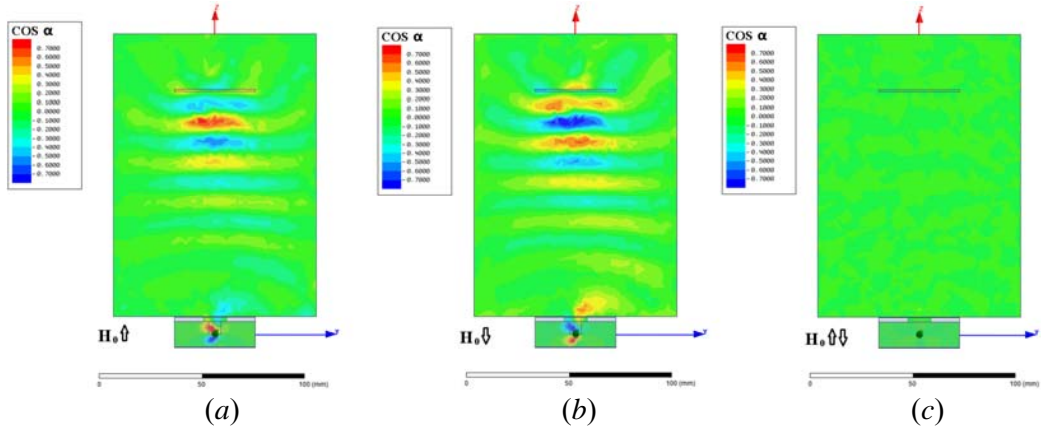

(*a*)            (*b*)            (*c*)

Fig. 6. A normalized helicity parameter in a vacuum far-field region calculated based on Eq. (8). (*a*), (*b*) At the frequency $f = 8.138$ GHz and opposite directions of a bias magnetic field; (*c*) at the frequency far from the MDM resonance, $f = 8.120$ GHz. Periodicity of the helicity-parameter striations is associated with the standing-wave behavior of regular EM waves in vacuum. Since a sign of a dynamical pseudoscalar field $\vartheta$ in Eq. (14) is correlated with a direction of a bias magnetic field applied to a ferrite disk, the striations change their signs, when a bias magnetic field is in an opposite direction. (*c*) At the frequency far from the MDM resonance, $f = 8.120$ GHz, one does not observe any field helicity ( $\cos \alpha = 0$ ).

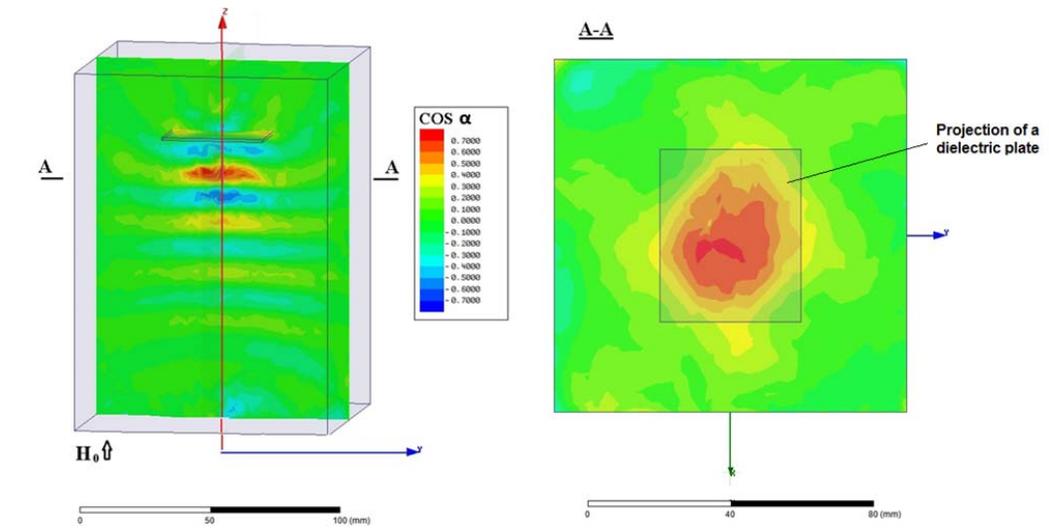

(a)

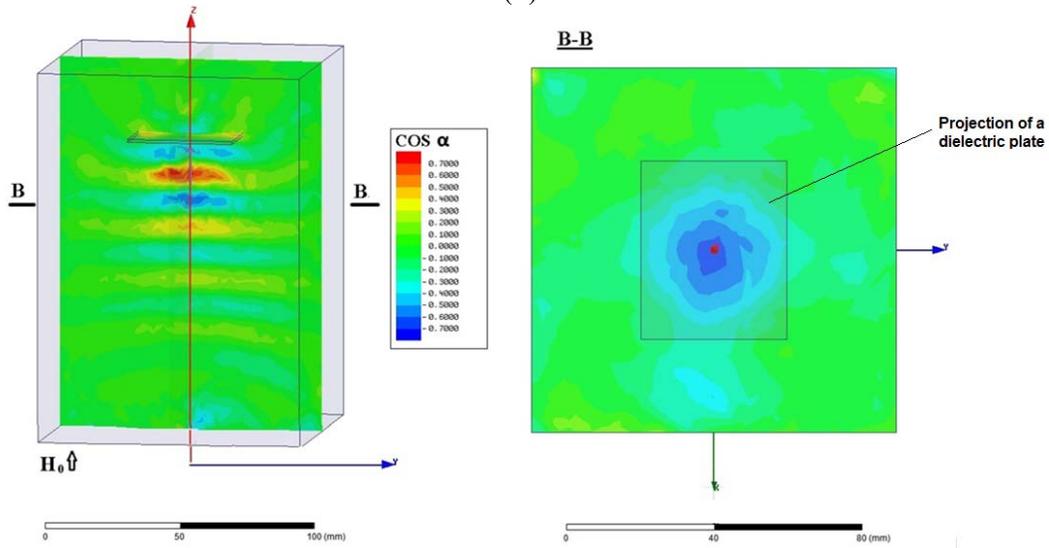

(b)

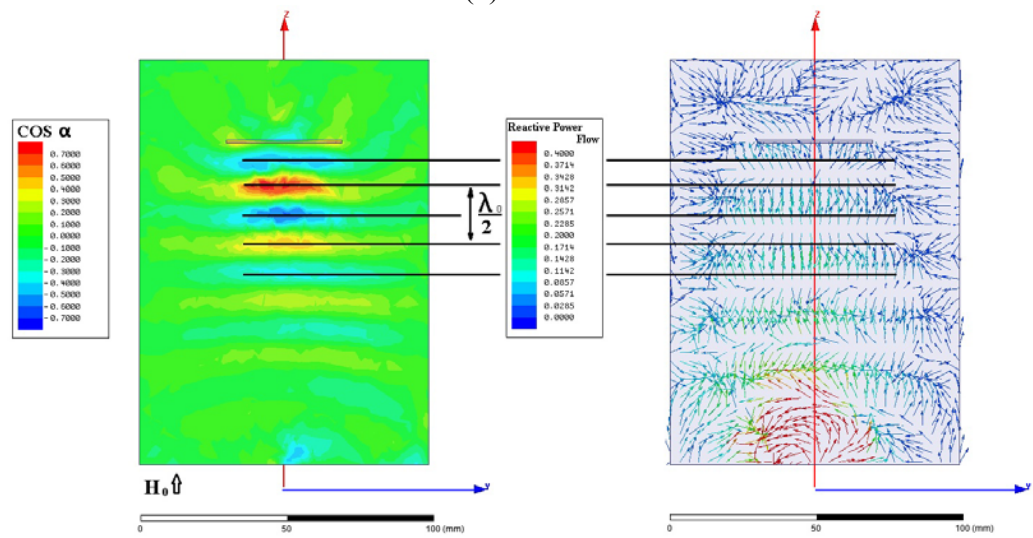

(c)

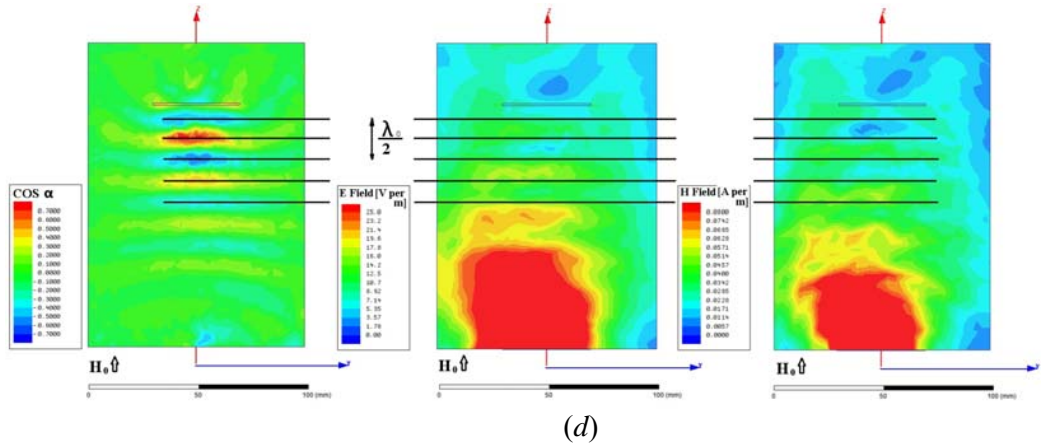

(d)

Fig. 7. A detailed analysis of the helicity parameter distribution in a vacuum far-field region at the frequency $f$ = 8.138GHz and a bias magnetic field directed along $z$ axis. (*a*), (*b*), The regions of localization of the helicity parameter on vacuum horizontal planes A-A and B-B for opposite directions of a bias magnetic field. (*c*) Along $z$-axis, the helicity parameter is correlated with the regions of localized reactive power flows. The maximums of the helicity-parameter modulus are at the same positions as the the regions of maximal gradient modulus of the reactive-power flow. (*d*), The maximums of the helicity-parameter modulus are at the regions of maximal gradient modulus of the field (both electric and magnetic) intensities. $\lambda_0$ is the electromagnetic wavelength in vacuum.

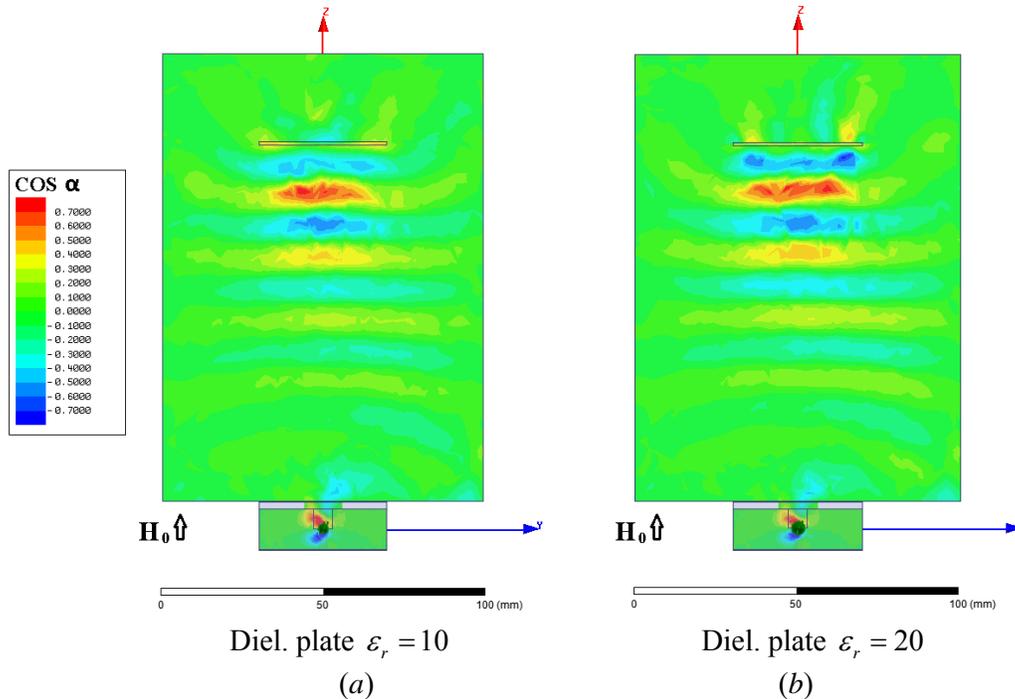

Diel. plate $\varepsilon_r = 10$        Diel. plate $\varepsilon_r = 20$

(*a*)        (*b*)

Fig. 8. The pictures of the helicity parameter distributions for two dielectric plates with different permittivity parameters and the same sizes. (*a*) $\varepsilon_r = 10$; (*b*) $\varepsilon_r = 20$.

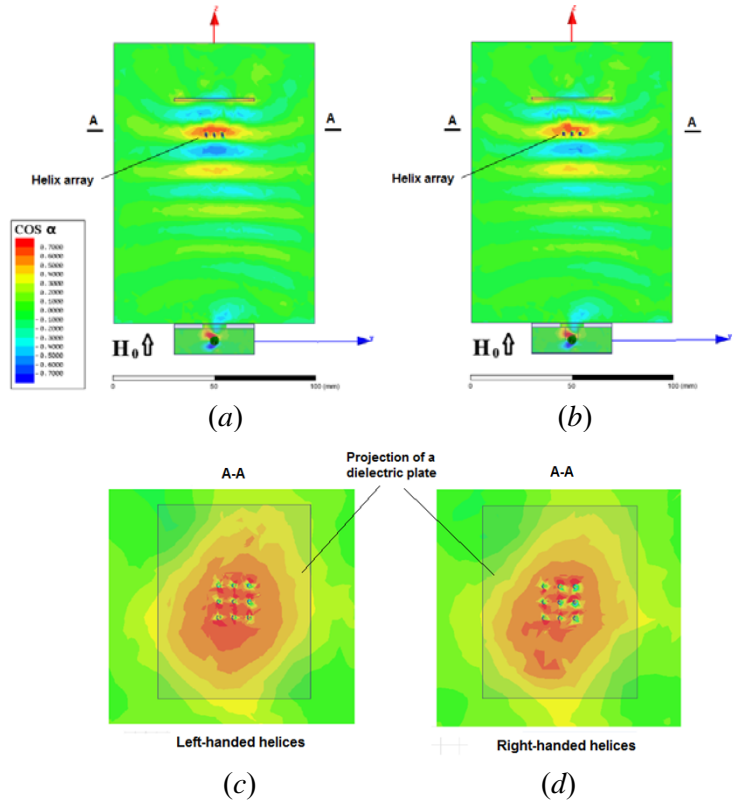

Fig. 9. An array of small metallic helices is placed in a space between an antenna and a dielectric plate ($\varepsilon_r = 10$). The helices do not change positions of the helicity-parameter striations. However, the regions inside every helix become green colored. It means that inside helices we have restoration of a regular EM-field behavior. (*a*), (*b*) side views; (*c*), (*d*) views on a vacuum horizontal plane; (*a*), (*c*) left-handed helices; (*b*), (*d*) right-handed helices.

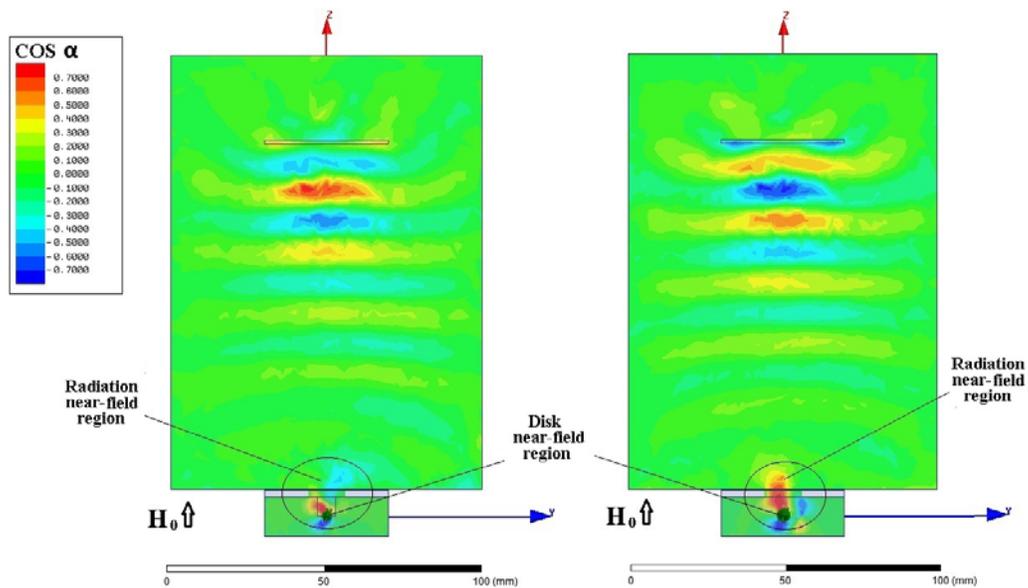

Fig. 10. Far-field helicity-factor distributions for the frequency *f* = 8.138GHz (the left slope of the resonance peak) and the frequency *f* = 8.140GHz (the right slope of the resonance peak). The selected regions show the helicity-factor distributions near a ferrite disk and near an antenna aperture. Near an antenna aperture the helicity-factor distributions are different for the frequencies *f* = 8.138GHz and *f* = 8.140GHz. Different helicity-factor distributions in the radiation near-field region are origins of different helicity-factor distributions in the far-field region. (*a*) Frequency *f* = 8.138GHz;   (*b*) frequency *f* = 8.140GHz.

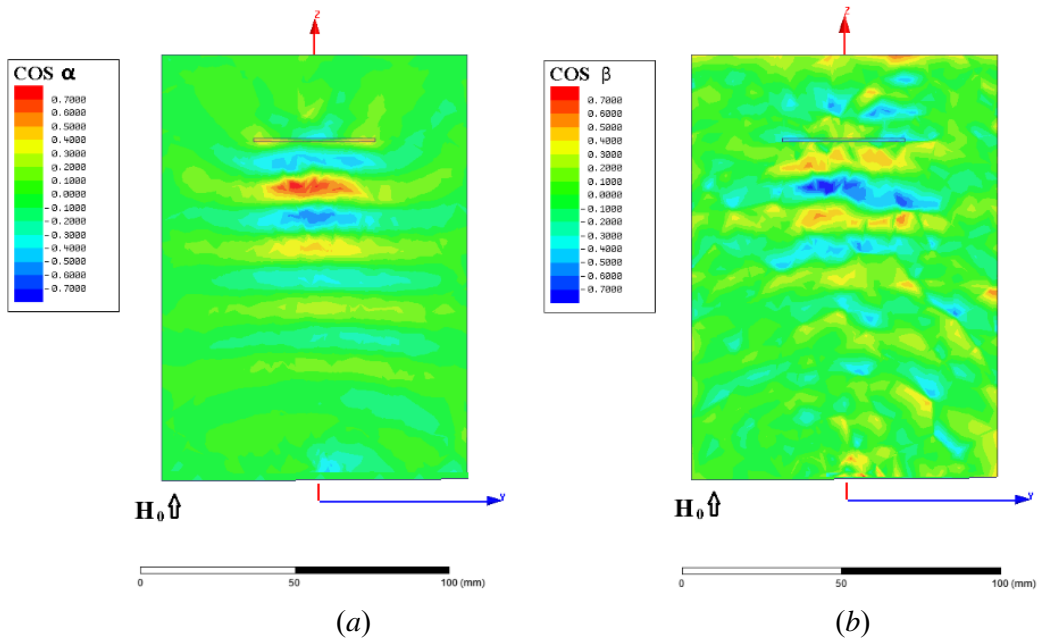

(*a*)                                                        (*b*)

Fig. 11. Normalized "electric" (*a*) and "magnetic" (*b*) helicity parameters in a vacuum far-field region at the frequency *f* = 8.138GHz, calculated based on the equations (34).

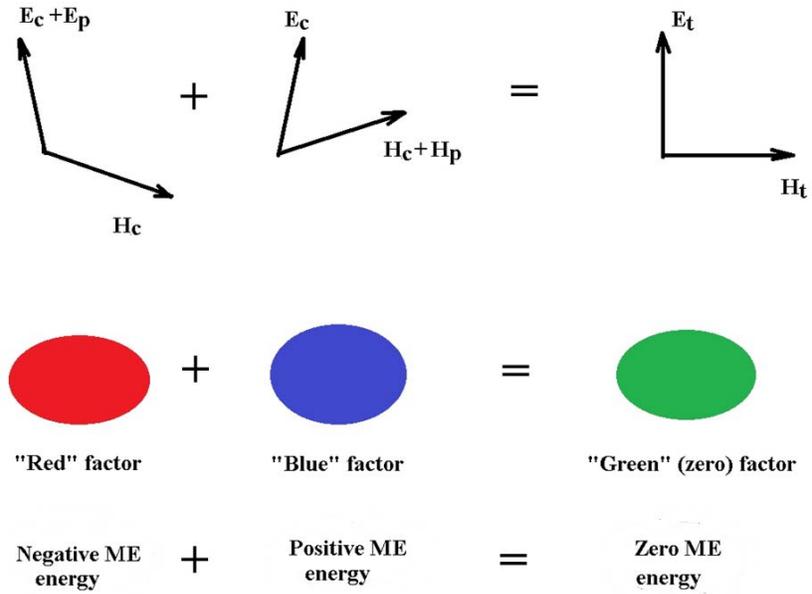

Fig. 12. In the far-field region, the turn of a total electric field $\vec{E}_t = \vec{E}_c + \vec{E}_p$ with respect to a curl magnetic field $\vec{H}_c$ is compensated by the turn of a total magnetic field $\vec{H}_t = \vec{H}_c + \vec{H}_p$ with respect to a curl electric field $\vec{E}_c$. This results in mutual perpendicularity between $\vec{E}_t$ and $\vec{H}_t$. In a local point in the far-field region, the "red" and "blue" helicity factors being superposed one to another, give the "green" (zero) helicity factor. In other words, the negative ME energy is compensated by the positive ME energy.